\documentclass[12pt]{iopart}

\usepackage{graphics} 
\begin{document}

\title[Universal optimal hole-doping concentration in single-layer HTSC]{Universal optimal hole-doping concentration in single-layer high-temperature cuprate superconductors}

\author{T Honma$^1$$^2$ and P H Hor$^1$}

\address{$^1$ Texas Center for Superconductivity and Department of Physics, \\University of Houston - Houston, TX. 77204-5002, USA}
\address{$^2$ Department of Physics, \\Asahikawa Medical College - Asahikawa, Hokkaido 078-8510, Japan}
\eads{\mailto{homma@asahikawa-med.ac.jp}, \mailto{phor@uh.edu}}

\begin{abstract}
We argue that in cuprate physics there are two types, hole content per CuO$_2$ plane ($P_{pl}$) and the corresponding hole content per unit volume ($P_{3D}$), of hole-doping concentrations for addressing physical properties that are two-dimensional (2D) and three-dimensional (3D) in nature, respectively. We find that superconducting transition temperature ($T_c$) varies systematically with $P_{3D}$ as a superconducting \textquotedblleft $dome$\textquotedblright\ with a universal optimal hole-doping concentration $P_{3D}^{opt.}$ = 1.6 $\times$ 10$^{21}$ cm$^{-3}$ for single-layer high temperature superconductors. We suggest that $P_{3D}^{opt.}$ determines the upper bound of the electronic energy of underdoped single-layer high-$T_c$ cuprates.

\end{abstract}

%\noindent{\sust}: Article preparation, IOP journals
\submitto{Supercond. Sci. Technol.}
\maketitle

\section{Introduction}
In high-temperature cuprate superconductors (HTS) the hole-doping concentration is the single most important physical parameter that dictates the cuprate physics. It has been a subject of extensive studies since the beginning of the HTS era until today. It is also evident from the existence of the robust generic electronic properties $vs.$ hole-doping concentration phase diagram \cite{bat96}. In spite of its fundamental important role in studying HTS, there is no consistent and simple method to determine hole-doping concentration for all HTS.

HTS consist of one, two and more conductive \textquotedblleft $CuO_2$ $plane$\textquotedblright\ layers sandwiched between blocks of insulating \textquotedblleft $charge$ $reservoir$\textquotedblright\ layers generally referred as single-, double- and multi-layer HTS, respectively. Cation-doping and/or oxygen-doping within the charge reservoir introduce holes into the CuO$_2$ plane. In general, the doped-holes are expressed as the {\textit{hole content per CuO}}$_2$ {\textit{plane}} ($P_{pl}$). Note that $P_{pl}$ so defined is intrinsically a two-dimensional (2D) quantity. In the cation doped HTS, the $P_{pl}$ can be directly determined from the doped cation content, such as $P_{pl}$ = $x$ in the Sr-doped La$_{2-x}$Sr$_x$CuO$_4$ (SrD-La214). However, in the oxygen-doped (OD-) or cation/oxygen co-doped (CD-) HTS, it is highly non-trivial to determine $P_{pl}$. Because, for La$_{2-x}$Sr$_x$CuO$_{4+\delta}$, $P_{pl}$ depends on the excess oxygen content ($\delta$), but the doping efficiency of oxygen atoms depends on the type of charge reservoir and/or $P_{pl}$ \cite{li96,tok88}.

Superconducting transition temperature ($T_c$) $vs.$ $P_{pl}$ for the SrD-La214 shows a dome-shaped curve with a maximum $T_c$ ($T_c^{max}$) of $\sim$ 37 K at $P_{pl}$ $\sim$ 0.16. Many other HTS also exhibited the similar dome-shaped curve, although the $T_c^{max}$ depends on the HTS materials. It was then conjectured that all HTS follow a universal dome-shaped curve with $T_c^{max}$ at $P_{pl}$ = 0.16 \cite{pre91}. Subsequently, the hole concentration ($P_{T_c}$), considered  to be identical to $P_{pl}$, was conveniently estimated by $T_c$($P_{T_c}$)-scale using eq.~(\ref{e.1}).

\begin{equation}
\label{e.1}
 \begin{array}{lr}
%\displaystyle T_{c}/T_{c}^{max} = 1 - 82.6 \Bigl(P_{T_c} - 0.16 \Bigr) ^2.
 \displaystyle \frac{T_{c}}{T_{c}^{max}} = 1 - 82.6 \Bigl(P_{T_c} - 0.16 \Bigr) ^2.
 \end{array}
\end{equation}

Most recently, based on the thermoelectric power at room temperature ($S^{290}$), a universal $S^{290}$($P_{pl}$)-scale (hereafter $P_{pl}$-scale) is constructed as new scale that is different from the $T_c$($P_{T_c}$)-scale (hereafter $P_{T_c}$-scale) \cite{hon04}. The distinct features of $P_{pl}$-scale are: (a.) pseudogap temperatures become universal to all HTS and depend only on $P_{pl}$, and (b.) $T_c^{max}$ is no longer universally pinned at $P_{pl}$ = 0.16, it depends on the specific material system of HTS \cite{hon04}. Therefore $P_{pl}$ can be regarded as a carrier scale dictated by the pseudogap energy scale. In contrast the $P_{T_c}$-scale is based on the energy scale of $T_c^{max}$. Since pseudogap phase is the precursor of high-$T_c$ superconductivity, it seems to be plausible that $P_{pl}$ is the proper carrier scale for both normal and superconducting properties of layered cuprates. If so, then we expect the $P_{pl}$-scale should recover the experimentally observed dome-shaped $T_c$($P_{T_c}$) curve. In this letter, we show that $P_{pl}$, with a straight forward extension of $P_{pl}$ to an effective three-dimensional (3D) hole concentration, can indeed describe the universal dome-shaped $T_c$($P_{T_c}$) curve.

We define a 3D hole concentration ($P_{3D}$) in terms of $P_{pl}$ in eq.~(\ref{e.2}). 

\begin{equation}
\label{e.2}
 \begin{array}{lr}
%\displaystyle P_{3D} \equiv P_{pl} \times \Bigl(n_{layer}/V_{u.c.} \Bigr ).
 \displaystyle P_{3D} \equiv P_{pl} \times \Bigl(\frac{N_l}{V_{u.c.}} \Bigr).
 \end{array}
\end{equation}
Here, $V_{u.c.}$ and $N_l$ are the unit cell volume and the number of CuO$_2$ plane per unit cell, respectively. Since $P_{3D}$ is defined on the universal 2D $P_{pl}$-scale, this definition has qualitatively taken into account the charge de-confinement effect of the holes in cuprates. Therefore $P_{3D}$ can be viewed as the \textquotedblleft $effective$\textquotedblright\ 3D hole-doping concentration even when holes are completely confined in $CuO_2$ $planes$. 

In this letter we make a clear distinction between $P_{pl}$, hole content per CuO$_2$ plane, and the corresponding effective 3D hole-doping concentration ($P_{3D}$) defined in eq.~(\ref{e.2}). We show that it is important to use the physically relevant hole-doping concentration in order to visualize the intrinsic and systematic doping behaviors of any physical property of HTS. Furthermore, we show that the $\textstyle \frac{T_c}{T_c^{max}}$ $vs.$ $P_{3D}$ exhibits a universal dome-shaped curve with the universal optimal hole-doping concentration $P_{3D}^{opt.}$ = 1.6 $\times$ 10$^{21}$ cm$^{-3}$ for single-layer HTS. We find that the $P_{T_c}$-scale is identical to the $P_{3D}$-scale and should be understood in the context of a normalized effective 3D carrier concentration. In this report we will focus only on the single-layer HTS. The extension of our definition of $P_{3D}$ for the multi-layer HTS with equivalent CuO$_2$ planes is straight forward.

\section{Experimental}
\begin{figure}
\includegraphics{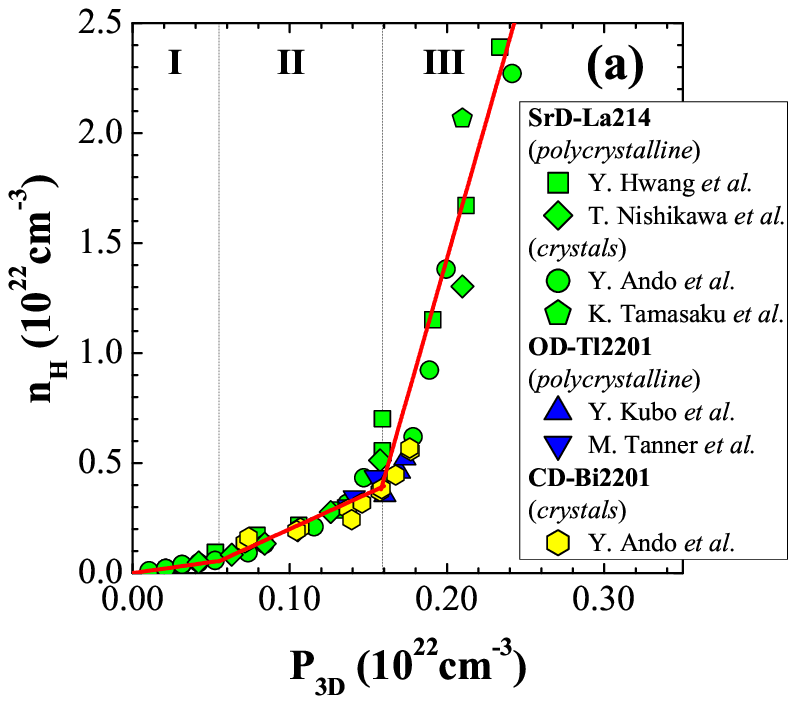}
\includegraphics{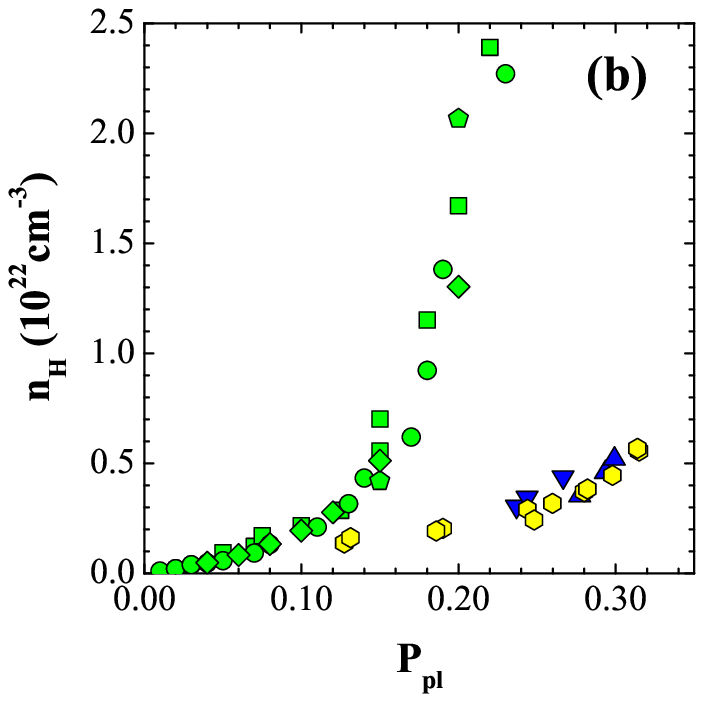}
\caption{(Colour online) (a) Hall number ($n_H$) as a function of $P_{3D}$ for the single-layer HTS. The slope of solid lines is 1 in regime-I, 3.2 in regime-II and 25 in regime-III. The data for SrD-La214 are extracted from ref.\ \cite{and01,tam94,nis95,hwa93}; CD-Bi2201 from ref.\ \cite{and00,and99,bal03}; OD-Tl2201 from ref.\ \cite{tan91,kub91}. The $P_{pl}$ of OD-Tl2201 by Kubo $et$ $al.$ \cite{kub91} is estimated from the $T_c$($P_{pl}$) curve. The dotted lines show $P_{3D}$ = 5.5$\times$10$^{20}$ cm$^{-3}$ and 1.6$\times$10$^{21}$ cm$^{-3}$. (b) Same data as in graph (a) are plotted as a function of $P_{pl}$.}
\label{f.1}
\end{figure}

$P_{pl}$ of data collected from the literatures \cite{obe92,rad94,and00,yam00,kom05,and99,bal03,and01,tam94,nis95,hwa93,tan91,uch96,hom04} are reliably determined by $P_{pl}$-scale. Some of the data plotted in Fig.~\ref{f.1} and \ref{f.3} are coming from the literatures \cite{wan00,pan03,hof98,kub91} where only $T_c$ is reported. Their $P_{pl}$ are estimated from the $T_c$($P_{pl}$) curve plotted in Fig.~\ref{f.2}(a). For the calculation of $P_{3D}$, we use the typical value of 190 \AA$^3$, 345 \AA$^3$, 355 \AA$^3$ and 143 \AA$^3$ for the unit cell volume of SrD-La214 with $N_l$ = 2 \cite{rad94}, Tl$_2$Ba$_2$CuO$_{6+\delta}$ (OD-Tl2201) with $N_l$ = 2 \cite{izu92}, Bi$_2$Sr$_{2-x}$La$_x$CuO$_{6+\delta}$ (CD-Bi2201) with $N_l$ = 2 \cite{tsv97}, and HgBa$_2$CuO$_{4+\delta}$ (OD-Hg1201) with $N_l$ = 1 \cite{wan93}, respectively.

\section{Results and discussion}
We start with a comparison of our scale to in-plane Hall number ($n_H$ = $\textstyle \frac{1}{|e|R_H}$), where $R_H$ is in-plane Hall coefficient and $|e|$ is electron charge, that estimates carrier number per unit $volume$.  In the single-layer SrD-La214 and OD-Tl2201, the $R_H$ of the polycrystalline samples is experimentally confirmed to be corresponding to the in-plane $R_H$ of the single crystals \cite{hwa93,man92}. In Fig.~\ref{f.1}(a), we plot the $n_H$ as a function of $P_{3D}$ for the single-layer HTS. For SrD-La214 \cite{and01,tam94,nis95,hwa93}, there are three linear $n_H$($P_{3D}$) regimes (regime-I, II and III). In regime-I for $P_{3D}$ $\le$ 5.5 $\times$ 10$^{20}$ cm$^{-3}$, $n_H$ is identical to $P_{3D}$. This is very encouraging and indicates that the $P_{3D}$ defined here is physically sound and quantitatively valid. At $P_{3D}$ = 5.5 $\times$ 10$^{20}$ cm$^{-3}$, the slope of linear $n_H$($P_{3D}$) suddenly changes from 1 to $\sim$3.2. In the regime-III for $P_{3D}$ $\ge$ 1.6 $\times $10$^{21}$ cm$^{-3}$, the linear $n_H$($P_{3D}$) changes slope to 25. The observed rapid increase in $R_H$ may related to the change in sign of $R_H$ observed in the overdoped SrD-La214 \cite{hwa93}. The three regimes clearly define three distinct electronic states of doped holes. The identical trend is also observed in CD-Bi2201 crystals \cite{and00,and99,bal03} and OD-Tl2201 ceramics \cite{obe92,tan91,kub91}. We need to emphasize that this systematic behavior is not governed by the $P_{pl}$, but by the $P_{3D}$. In Fig.~\ref{f.1}(b) we plot the same data set of $n_H$ as a function of $P_{pl}$. The $n_H$ for CD-Bi2201 and OD-Tl2201 do not follow that of SrD-La214, and the three physically distinct regimes can not be resolved.

\begin{figure}
\includegraphics{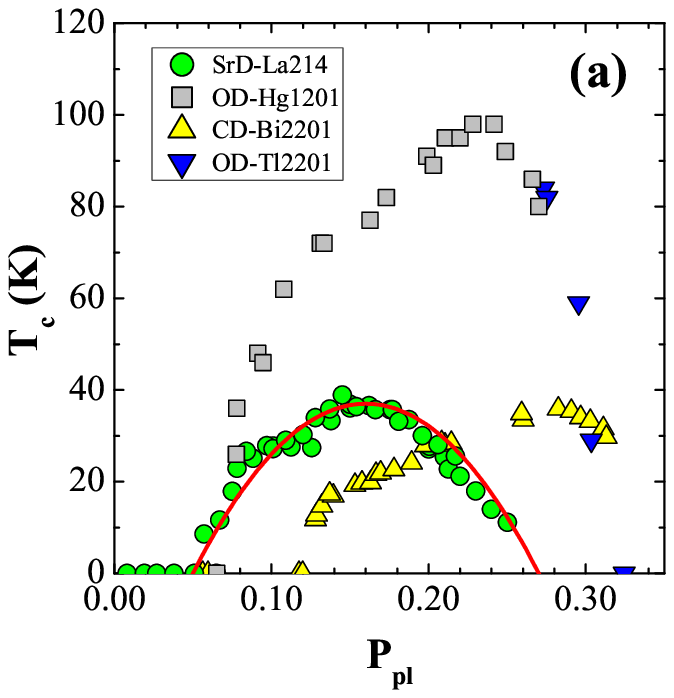}
\includegraphics{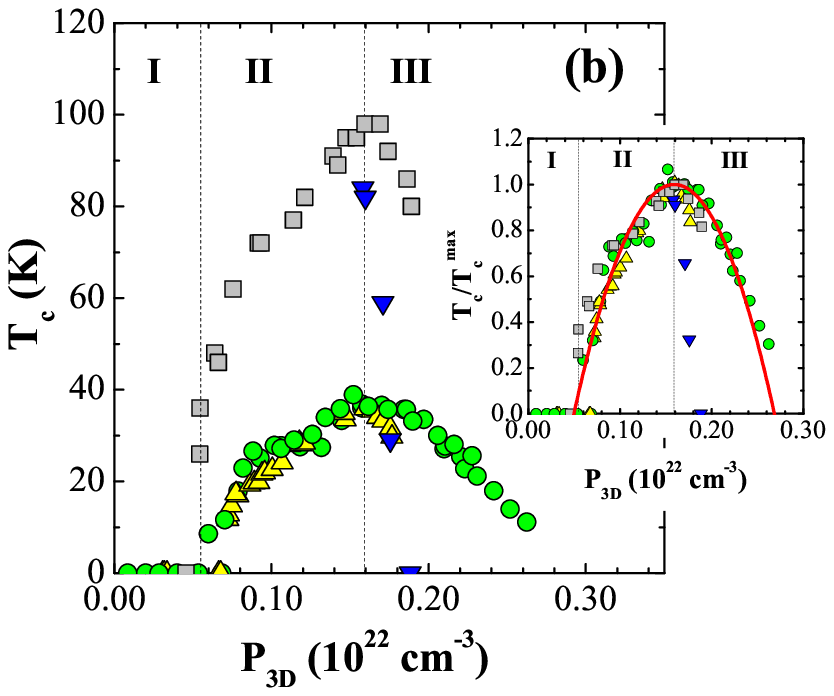}
\caption{(Colour online) (a) Superconducting transition temperature ($T_c$) as a function of $P_{pl}$ for the single-layer HTS. The data for SrD-La214 are extracted from ref.\ \cite{rad94,kom05}. The data set for OD-Hg1201, CD-Bi2201 and OD-Tl2201 are extracted from ref.\ \cite{yam00},\ \cite{and00}, and\ \cite{obe92}, respectively. The solid line shows $T_c$ curve ($T_c^{max}$ = 37 K) reproduced by using eq.~(\ref{e.1}). (b) Same data as in graph (a) are plotted as a function of $P_{3D}$. The dotted lines show $P_{3D}$ = 5.5 $\times$ 10$^{20}$ cm$^{-3}$ and 1.6 $\times$ 10$^{21}$ cm$^{-3}$. The inset shows the $\textstyle \frac{T_c}{T_c^{max}}$ $vs.$ $P_{3D}$. The solid line shows $\textstyle \frac{T_c}{T_c^{max}}$ curve reproduced by using eq.~(\ref{e.3}).}
\label{f.2}
\end{figure}

Since it is well known that the electronic anisotropy decreases with doping in all HTS, the above observations suggest that there are three distinct confinement regimes of doped holes. In regime-I where $n_H$ is the same as $P_{3D}$ indicates doped holes are completely confined within the CuO$_2$ plane. For regimes-II and -III where $n_H$ = $\alpha$$P_{3D}$ with a non-unity slope, $\alpha$ $\sim$ 3.2 and 25, respectively, each represents a distinct de-confinement regime of doped-hole states. Noted that the change of slope between two regimes is quite abrupt, more like a phase transition than some kind of crossover behaviors. The critical $P_{3D}$ that define the boundary between regime-I and -II and the boundary between regime-II and -III actually correspond to the crtitical $P_{3D}$ for the insulator-superconductor transition ($P_{3D}^c$) and $P_{3D}^{opt.}$, respectively. The physical meaning of each regime will become clear in our discussion of Fig.~\ref{f.2}(b).

In Fig.~\ref{f.2}(a), we plot $T_c$ as a function of $P_{pl}$. Here, we use $T_c$ value reported in the literature \cite{obe92,rad94,and00,yam00,kom05}, irrespective of how it was defined. The optimal doping level strongly depends on the systems. Noted that the optimal doping levels for OD-Hg1201 and CD-Bi2201 are at $P_{pl}$ $\sim$ 0.23 and 0.28, respectively, although that for SrD-La214 is $\sim$ 0.16. Thus, the optimal doping level for the single-layer HTS is not universally equal to 0.16, as reported in some earlier studies \cite{hon04,mar02}. However, the $P_{T_c}$-scale has been used successfully to discuss various characteristic properties of HTS \cite{tal01}. We show that the $P_{T_c}$ is actually related to $P_{3D}$. 

To see that $P_{3D}$ has real physical consequence we replot, in Fig.~\ref{f.2}(b), $T_c$ as a function of $P_{3D}$ using the same data set of Fig.~\ref{f.2}(a). The superconductivity appears at $\sim$ 5.5 $\times$ 10$^{20}$ cm$^{-3}$. The $T_c^{max}$ universally appears at $\sim$ 1.6 $\times$ 10$^{21}$ cm$^{-3}$. The inset shows the $\textstyle \frac{T_c}{T_c^{max}}$ $vs.$ $P_{3D}$. The $\textstyle \frac{T_c}{T_c^{max}}$ for SrD-La214, OD-Hg1201 and CD-Bi2201 follow the same dome-shaped curve. Now we can pin down the absolute value of 3D optimal hole-doping concentration in eq.~(\ref{e.3}) :
\begin{equation}
\label{e.3}
 \begin{array}{lr}
%  \displaystyle T_{c}/T_{c}^{max} = 1 -83.64 \Bigl(P_{3D} \times 10^{-22} - 0.159 \Bigr) ^2, 
  \displaystyle \frac{T_{c}}{T_{c}^{max}} = 1 -83.64 \Bigl(P_{3D} \times 10^{-22} - 0.159 \Bigr) ^2, 
 \end{array}
\end{equation}
where the unit of $P_{3D}$ is \textquotedblleft\ $cm^{-3}$ \textquotedblright. It is clear that the $P_{T_c}$ determined in eq.~(\ref{e.1}) is not planar hole-doping concentration but physically identical to our defined $P_{3D}$. Therefore, we can understand why $P_{T_c}$-scale worked in the earlier doping-dependence studies. However, we need to emphasize that $P_{T_c}$-scale is the proper carrier scale for 3D \textquotedblleft bulk\textquotedblright\ cuprate properties. The $T_c$ of underedoped  OD-Hg1201 may seem slightly to be higher than $T_c$ calculated from eq.~(\ref{e.3}). OD-Hg1201 is the pure oxygen-doped compound, although CD-Bi2201 is the cation/oxygen co-doped compound. Accordingly, the oxygen atom within the charge reservoir is more mobile or softer in OD-Hg1201 \cite{sad00,lor02}. The slightly higher $T_c$ could be due to the change in $P_{pl}$ by the thermal oxygen re-arrangement \cite{lor02}.

\begin{figure}
\includegraphics{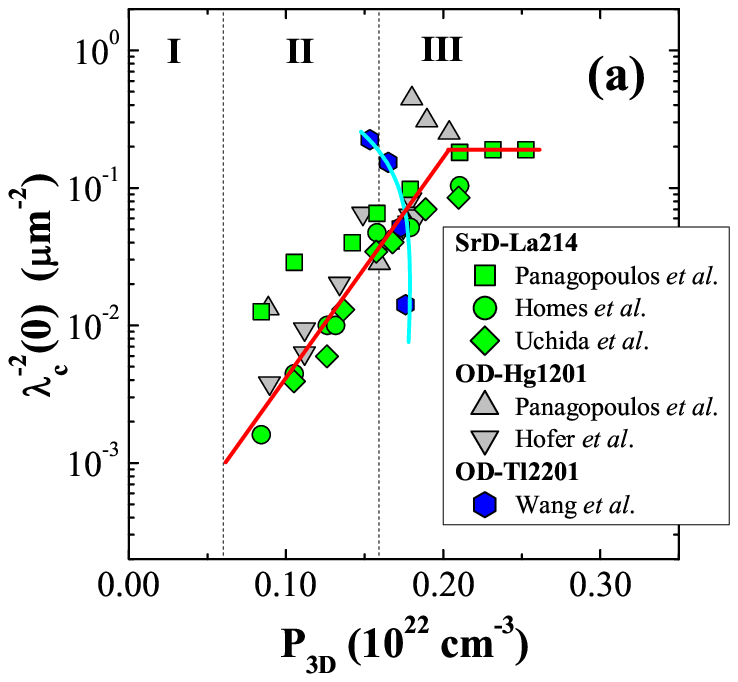}
\includegraphics{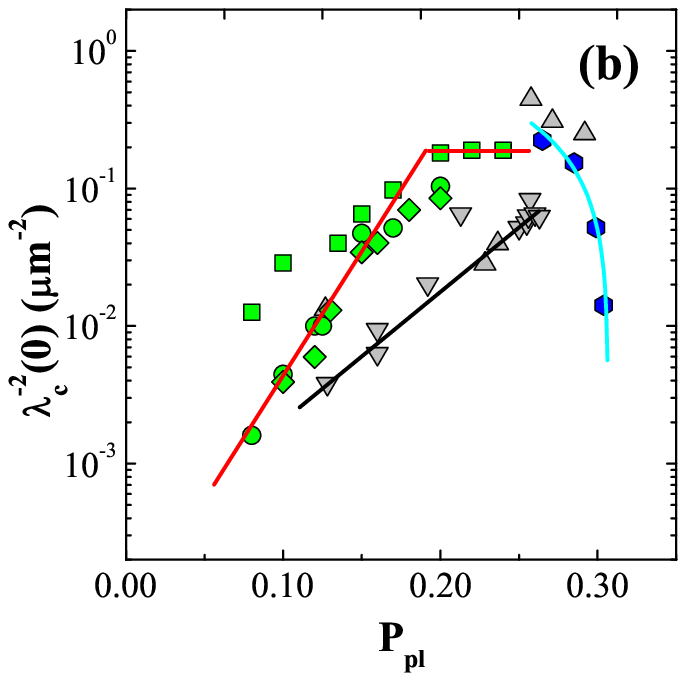}
\caption{(Colour online) (a) $c$-axis logarismic inverse squared penetration depth ($\lambda_c^{-2}$(0)) as a function of $P_{3D}$. The data for SrD-La214 are extracted from ref.\ \cite{uch96,hom04,pan03}, OD-Hg1201 from ref.\ \cite{pan03,hof98} and OD-Tl2201 from ref.\ \cite{wan00}. The dotted lines show $P_{3D}$ = 5.5$\times$10$^{20}$ cm$^{-3}$ and 1.6$\times$10$^{21}$ cm$^{-3}$. (b) Same data as in graph (a) are plotted as a function of $P_{pl}$. The solid lines are guide for the eyes.}
\label{f.3}
\end{figure}

To further demonstrate our point that we should use the hole-doping concentration of the correct dimensionality consistent with the very nature of the physical properties under study, we consider the inverse square of the $c$-axis penetration depth ($\lambda_c^{-2}$(0)). The $c$-axis superfluid density ($\lambda_c^{-2}$(0)), independent of the $c$-axis coupling mechanism, is intrinsically a 3D property of the HTS. In Fig.~\ref{f.3}(a), we plot $\lambda_c^{-2}$(0) on the logarithmic scale as a function of $P_{3D}$ \cite{uch96,hom04,wan00,pan03,hof98}. The $\lambda_c^{-2}$(0) increases with doping until 2 $\times$ 10$^{21}$ cm$^{-3}$, higher than $P_{3D}^{opt.}$, and either decreases or remains at a constant level beyond. In Fig.~\ref{f.3}(b), we plot the same data set of $\lambda_c^{-2}$(0) as a function of $P_{pl}$. In the underdoped side, the trend in $\lambda_c^{-2}$(0) for OD-Hg1201 is different from that for SrD-La214. No systematic $\lambda_c^{-2}$(0) $vs.$ $P_{pl}$ behaviors can be observed. Thus, the $\lambda_c^{-2}$(0) is not governed by the $P_{pl}$ but by the $P_{3D}$. Accordingly, the $\lambda_c^{-2}$(0), independent of the $c$-axis coupling mechanism, is intrinsically a 3D property of the HTS.  There are many studies based on the $P_{T_c}$-scale \cite{tal01} and, to put them into proper prospective, these results should be understood in the context of $P_{3D}$.

The existence of a universal $P_{3D}^{opt.}$ has an interesting implication to the stability of underdoped electronic states, namely, it is the upper bound of the electronic energy of single-layer HTS. Therefore there is a common physical origin, dictated by $P_{3D}^{opt}$, for the transition from underdoped to overdoped regimes. Note that $P_{3D}^{opt.}$ is an effective \textquotedblleft 3D\textquotedblright\ hole concentration  that is intrinsically consistent with the dominance of Fermi liquid state in 3D electronic systems and the experimental observations that Fermi liquid behavior emerges beyond optimal doping concentration \cite{tak05}. Using the simple free electron model with hole concentration equal to $P_{3D}^{opt.}$ \cite{kit05}, the corresponding Fermi energy is $\sim$ 0.5 eV, close to the binding energy of the lower Hubbard band ($\sim$ 0.4 to 0.6 eV) of layered cuprates \cite{yos03}. More specifically, as suggested in ref. \cite{hor02}, the $\sim$ 0.5 eV is the energy required to promoting a hole from $d_{x^2-y^2}$ to $d_{3Z^2-r^2}$ to form $d-d^*$ exciton, another plausible channel for hole de-confinement.

\section{Summary}
We have shown that for HTS there are two types of hole-doping concentration depending on the dimensionality, that is, the effective 3D hole-doping concentration $P_{3D}$ and hole content per Cu-O$_2$ plane $P_{pl}$ defined in ref.\cite{hon04}. Combining these two we have a complete working scale to address various physical properties for all HTS. Any 2D/in-plane property should be plotted as a function of $P_{pl}$ and any 3D/out-of-plane property as a function of $P_{3D}$. Indeed, we see that $\lambda_c^{-2}$(0) and the magnitude of $T_c$ are governed by $P_{3D}$, while pseudogap physics is described by $P_{pl}$ \cite{hon04}. While the $T_c^{max}$ is different for different single-layer HTS, we observed a universal $P_{3D}^{opt.}$ = 1.6 $\times$ 10$^{21}$ cm$^{-3}$ for single-layer HTS that determines the stability limit of the electronic states of the underdoped regime. The optimal superconducting transition temperature for a specific material system will depend on another material specific energy-scale, such as the inter-plane block spin coupling suggested in ref. \cite{kim06}, along the $c$-axis. 

\ack{Acknowledgments}
One of us (TH) would like to thank Professor M Tanimoto of Asahikawa Medical College for providing the administrative convincement for this study. This work was supported by the State of Texas through the Texas Center for Superconductivity at the University of Houston.

\end{document}